\documentclass[5p]{elsarticle}
\newcounter{bla}

\usepackage{amsmath}
\usepackage{amssymb}
\usepackage{colordvi}
\usepackage{graphicx}
\usepackage{bm}
\bmdefine{\bx}{x}

\journal{Computer Physics Communications }

\begin{document}
\begin{frontmatter}

\title{Multicanonical Sampling of Rare Trajectories 
 in Chaotic Dynamical Systems}
\author[a]{Akimasa Kitajima\corref{author}}
\author[b]{Yukito Iba}
\ead{kitajima@cp.cmc.osaka-u.ac.jp, iba@ism.ac.jp}

\cortext[author]{Corresponding author}
\address[a]{Graduate School of Science and
 Cybermedia Center, Osaka University,
Toyonaka, Osaka 560-0043, Japan}
\address[b]{The Institute of Statistical Mathematics,
10-3 Midorimachi, Tachikawa, Tokyo 190-8562, Japan
}

\begin{abstract}
In chaotic dynamical systems, a number of rare trajectories
with low level of chaoticity are embedded in chaotic sea,
while extraordinary unstable trajectories can exist 
even in weakly chaotic regions. In this study,
a quantitative method for searching these rare trajectories
is developed; the method is based on multicanonical
Monte Carlo  and can estimate
the probability of initial conditions that lead to trajectory fragments
of a given level of chaoticity. The proposed method is successfully
tested with four-dimensional coupled standard maps, where
probabilities as small as $10^{-14}$  are estimated.


\end{abstract}

\end{frontmatter}


\sloppy

\section{Introduction}

In chaotic dynamical systems, a number of rare trajectories
with low level of chaoticity are embedded in chaotic sea, which
show apparently small values of Lyapunov exponents.
Also, there can be extraordinary unstable trajectories with 
apparently large values of Lyapunov exponent even 
in weakly chaotic systems.
  
Numerical search for such rare trajectories
and  quantitative estimation of their probabilities are
important for the understanding of these dynamical systems.
Even when Lyapunov exponent is converged 
to a unique value almost everywhere in a chaotic sea, 
the probability of finite-length trajectories of a given
chaoticity reflects transient behavior
of the system and  hence  provides information on fine structures 
in the state space~\cite{BeckSchloegl199501}.

In this paper, we develop a method for calculating
probabilities of initial
conditions that lead to trajectory fragments
of a given level of chaoticity.
This method is based on the
multicanonical Monte Carlo algorithm~\cite{
berg1992multicanonical},
which is a version of dynamic Monte Carlo (Markov chain
Monte Carlo)~\cite{metropolis1953equation, 
landau2009guide}
and provides a powerful tool for calculating
small probabilities under a  given probability measure.
The proposed method is tested with four-dimensional 
coupled standard maps.  It is shown that the 
method  can deal with rare events
of very small probabilities such as $\sim 10^{-14}$.

The method proposed in this paper can be regarded as  
a variant of the method recently 
introduced in Yanagita and Iba~\cite{yanagita2009exploration};
the proposed method uses the multicanonical
Monte Carlo instead of the replica exchange
Monte Carlo used in~\cite{yanagita2009exploration}.  
A novelty in this paper is that 
probabilities of initial conditions
that generate rare trajectories are successfully
estimated, while no quantitative result 
is obtained in the previous paper~\cite{yanagita2009exploration}.
Also, the multicanonical algorithm has
advantages over the replica exchange Monte Carlo;
it enables direct computation of the desired
probabilities and is efficient in cases with first order
transitions~\cite{berg1992multicanonical}.

Our studies are partly motivated by the studies~\cite{Tailleur07}, 
where rare structures with high or low
chaoticity are explored by simulating fictitious particles that 
are moved, split, and  systematically erased; their method can be
regarded as tracking rare ``pseudo-orbits'' in a parallel manner.
The method proposed in this paper samples initial conditions 
and seems complementary to their approach. 
There are also
references~\cite{Sasa06, Takeuchi07, Pratt86, Cho94} 
that discuss sampling of unstable periodic 
orbits in chaotic systems by dynamic Monte Carlo or
related methods.
None of these studies, however,  seems
to compute probabilities of rare trajectory fragments
using the multicanonical algorithm.

\section{Algorithm} \label{algorithm}

\subsection{Forgetting Time as a Measure of Chaoticity}

Let us consider a deterministic dynamical system
\mbox{$x(t+1)=\psi(x(t))$} with discrete time $t=1, 2, \ldots$
and assume a trajectory $x(t)$  originated from an initial
condition $x(0)=x_0$.  Specify a  measure $f$ for the chaoticity
of the trajectory $x(t)$  and consider it  
as an integer-valued function $f(x_0)$ of the initial condition $x_0$;
for a real-valued $f$, appropriate discretization of
its value is assumed.
Then, our interest is in estimating the probability
$$
P(\tilde{f})= \frac{1}{D} \int  \delta(f(x_0)-\tilde{f}) dx_0 
$$
where $\delta$ is defined as $\delta(s)=1$ if $s=0$;
$\delta(s)=0$ otherwise. $dx_0$ is uniform measure 
on the space of initial conditions and $D$ is the volume
of the entire space.

An important point is the choice of $f(x_0)$; assuming
finite precision arithmetic of computers, it should be
something like ``approximate Lyapunov exponent'' of
a finite piece of the trajectory starting from $x_0$.

A possible way is to fix the length $T$ of the trajectories
and define $f(x_0)$ as the largest eigenvalue $\lambda_T(x_0)$
of the  product 
$\prod_{t=1}^T \mathbf{J}(x(t))$
of the Jacobian matrices $\mathbf{J}(x(t))$
along the piece of the trajectory $x(t)$ starting from $x_0$. 
Here $(i, j)$-component of 
$\mathbf{J}(x(t))$ is given by
$
\mathbf{J}_{ij}(x(t))= \left. \frac{\partial \psi_i (x)}{\partial x_j} \right |_{x=x(t)}, 
$
where $\psi_i$ and $x_i$ are $i$th component
of the function $\psi$ and its argument, respectively. 
This way of choosing $f(x_0)$ is, however, not suited
for numerical computation when $\lambda_T(x_0)$ is
strongly dependent on $x_0$. When $\lambda_T(x_0)$  is large
we should choose small $T$ to control the effect 
of  round-off error. On the other hand, larger $T$ is desirable
in the region of small $\lambda_T(x_0)$ to filter out trajectories
with low chaoticity. Thus, it is difficult to choose the value
of $T$ adequate for all values of $\lambda_T(x_0)$.

Our solution is to reverse the idea and 
use the minimum value $T_\epsilon(x_0)$ of $T$ that
satisfies $\lambda_T(x_0)\, \epsilon > 1$ as a
measure of chaoticity, 
where $\epsilon$ is a small constant beyond the machine
epsilon;  larger value of $T_\epsilon(x_0)$ means
larger  number of time-steps required to
forget the initial condition and  implies lower chaoticity.
The switching from $f(x_0)=\lambda_T(x_0)$
to ``forgetting time'' $f(x_0)=T_\epsilon(x_0)$  provides
a computationally stable criterion of chaoticity,
because by definition the latter  is 
insensitive to round-off error.

\subsection{Multicanonical Algorithm}

We introduce the  multicanonical algorithm to 
estimate the probability $P(f)$; the algorithm 
consists of two stages, training and measurement.
In the training phase,  we  construct
the approximate probability $\tilde{P}(f)$
step-by-step through dynamic Monte Carlo simulations.
At each step, we perform Monte Carlo simulation 
with the weight function $1/\tilde{P}(f(x_0))$ using 
the current estimate of  $\tilde{P}(f)$.
If $\tilde{P}(f)$ is a good approximation to $P(f)$  
in a prescribed interval of $f$,
 the histogram $h(f)$ is  almost flat in the interval, because 
$$
h(f) \propto \frac{1}{D} \int \delta(f(x_0)-f) \frac{1}{\tilde{P}(f(x_0))} dx_0=
\frac{P(f)}{\tilde{P}(f)} \simeq 1.
$$
If $h(f)$ is not sufficiently flat, we modify $\tilde{P}(f)$ until
an almost flat histogram of $f$ is obtained. 
Several methods are proposed for efficient tuning of 
the weight in the training phase; in this study, we
use a method due to Wang and Landau~\cite{
wang2001determining}.
After we obtain a good approximation of $P(f)$
we enter the measurement phase and perform a long run of 
simulation with a fixed weight $1/\tilde{P}(f(x_0))$.
Then, the final estimate $P^*(f)$ of $P(f)$ 
is obtained as $P^*(f) \propto h(f)\tilde{P}(f) $.

The key to the implementation of the algorithm is that 
the weight  $1/\tilde{P}(f(x_0))$  
of $x_0$ is expressed with
a composite function of a univariate function $\tilde{P}(f)$
and  a known function $f(x_0)$; it enables easy adaptation
of the weight using outputs of the simulations. 
On the other hand,  the multicanonical algorithm
enjoys  fast mixing of the Markov chain;
sampling in a wide range of $f$
realizes a kind of ``annealing'' effect and the convergence
becomes much better compared with the cases where 
only the tail regions are sampled.  

So far we discuss the generic algorithm. 
To implement the multicanonical algorithm in the
present case, where $f(x_0)=T_\epsilon(x_0)$, 
we should specify dynamic Monte
Carlo algorithm for the sampling with the
weight $1/\tilde{P}(T_\epsilon(x_0))$. 
Here we use the Metropolis 
algorithm~\cite{metropolis1953equation, 
landau2009guide}, in which
a candidate $x_0^{new}$ is generated 
using a hierarchical proposal distribution
used in~\cite{yanagita2009exploration}. The candidate $x_0^{new}$
is accepted if and only if a uniform random number
$r \in [0,1)$ satisfies \mbox{$r<\frac{\tilde{P}(T_\epsilon(x^{old}_0))
}{\tilde{P}(T_\epsilon(x^{new}_0))}$}, where $x^{old}_0$ 
is the current value of $x_0$; to compute 
$T_\epsilon(x^{new}_0)$
at each step of the Metropolis algorithm, we simulate
the trajectory from the initial condition $x^{new}_ 0$
until \mbox{$\lambda_T(x^{new}_0) \, \epsilon > 1$}.

\subsection{Computation of the Largest Eigenvalue} \label{subsec_power}

To implement the proposed algorithm for high-dimensional
dynamical systems, we should compute the largest eigenvalue $\lambda$
of the matrix $\mathbf{J}_T=\prod_{t=1}^T \mathbf{J}(x(t))$ efficiently.
In this study, we approximate it  using the power method as 
$\lambda \simeq ||\mathbf{J}_T^m \xi||$, where
$\xi$ is  a constant vector of unit length $||\xi||=1$. 
We found that $m=1$
often gives a good approximation.

\section{Numerical Experiments} \label{experiments}

\subsection{Searching Low Chaoticity}

Let us consider four-dimensional coupled standard maps
\begin{equation} \label{eq:map} 
\begin{split}
  u_{n+1} &= u_n-\frac{K}{2\pi}\sin(2\pi v_n)+\frac{b}{2\pi}\sin(2\pi(v_n+y_n)), \\
  v_{n+1} &= v_n+u_{n+1},  \\
  x_{n+1} &= x_n-\frac{K}{2\pi}\sin(2\pi y_n)+\frac{b}{2\pi}\sin(2\pi(v_n+y_n)), \\
  y_{n+1} &= y_n+x_{n+1},
 \end{split}
\end{equation}
which is a well-studied example of volume preserving maps. 

\begin{figure}[tbp]
\centering
\includegraphics[width=0.4\textwidth]{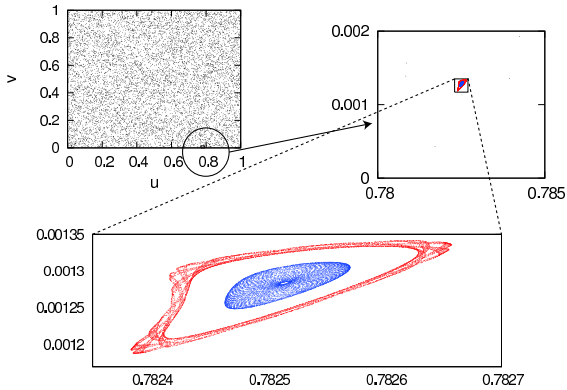}
\caption{A pair of tiny tori in a chaotic sea found 
by the proposed method. Projections on 
the $(u_n,v_n)$-plane are shown.
Enlargement of a tiny area 
in the small circle in the left 
panel is given in the right panel;  further enlargement
is given in the lower panel. $K=7.8$ and $b=0.001$.}
\label{fig_torus}
\end{figure}

First, we test the efficiency of the proposed algorithm
to find tiny tori in chaotic sea. 
In Fig.~\ref{fig_torus}, a pair of tori in chaos found by the
algorithm is plotted on the $(u_n,v_n)$-plane;
values of parameters $K=7.8$ and $b=0.001$ are chosen that
most of the state space is covered by a chaotic sea.
The threshold $\epsilon$ is $2^{-43}$.
The values of 
 $T_\epsilon$ corresponding
to the initial conditions
that lead to these tori are larger than $199$ and the probabilities
$P(T_\epsilon)$ are as small as $ 10^{-12}$.  The number
of initial conditions tested in the proposed method
is about $4\times 10^{9}$.

\subsection{Probabilities $P(T_\epsilon)$ }

\begin{figure}[th]
\centering
\includegraphics[width=0.33\textwidth]{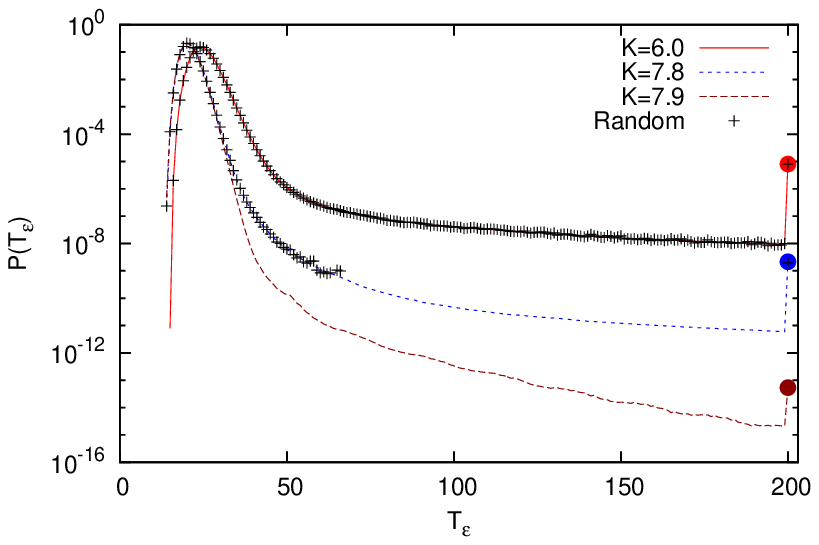}
\includegraphics[width=0.33\textwidth]{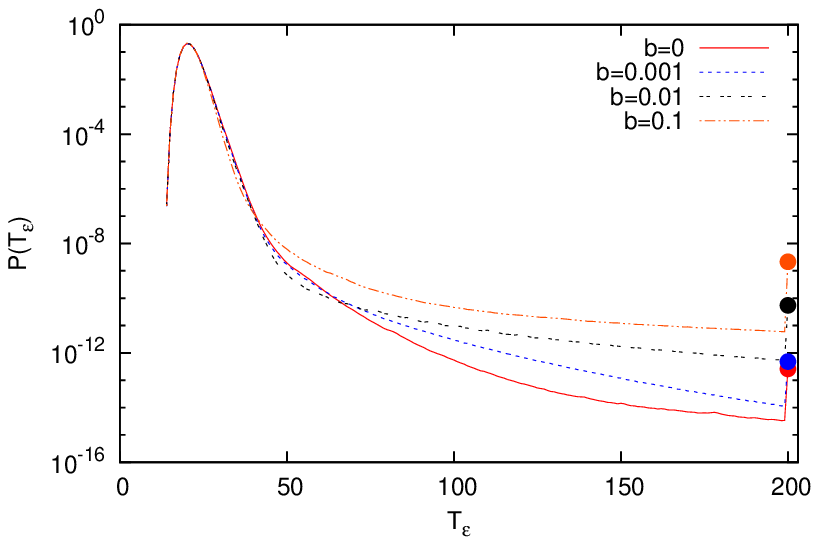}
\caption{
The probability $P(T_\epsilon)$ is plotted 
as a function of $K$(upper panel, $b=0.1$)
and $b$(lower panel, $K=7.8$), where $\epsilon=2^{-43}$.
The results of naive random sampling
are  shown by the symbol $+$, only when 
more than $10$ samples are available. 
The symbols {$\bullet$} 
at the right edge of the plot indicate values of the
probability $P(T_\epsilon \ge 200)$. 
}
\label{fig_prob1}
\end{figure}

An advantage of the multicanonical algorithm is
that it computes probabilities $P(T_\epsilon)$ 
of $T_\epsilon$ under uniform
sampling of initial conditions.
The probability of trajectory fragments that 
stays long time near tiny tori and/or cantori
will reflect fine structures
of phase space, which can be quantified by $P(T_\epsilon)$.
In Fig.~\ref{fig_prob1}, 
$P(T_\epsilon)$ for the model~\eqref{eq:map}
is plotted. The result indicates
that,  in this parameter range, the volume of
small regular regions decrease under
increasing $K$, while increase under increasing $b$.

The number of initial conditions tested in 
the proposed algorithm is $3\times 10^9 \sim 3\times 10^{10}$ 
for each curve including training phase and 
rejected candidates;  the corresponding 
computational times are
$17 \sim 125$ hours
using a single core of 
Intel Xeon X5365, which is a quad-core CPU.
These results show that
the proposed method can compute probabilities
down to $\sim 10^{-14}$ within reasonable computational times.

For the comparisons, results of naive random sampling
are also shown in Fig.~\ref{fig_prob1}. 
For $(K, b)=(6.0,0 .1)$,
there is an overall agreement of the outputs following
from the two sampling approaches.
For $(K, b)=(7.8, 0.1)$,  both results are also consistent, 
but naive random sampling
with comparable computational efforts 
($1.4\times 10^{10}$ initial
configurations) gives meaningful results
only in a high probability region, as seen
in the upper panel of Fig.~\ref{fig_prob1}.

In these examples, we set the number $m$ 
of iteration defined
in Sec.~\ref{subsec_power} to unity;
$m=2$ and $3$ are also tested and
slight differences are found in some cases.

\section{Summary and Discussion} \label{discussion}
 
A quantitative method 
based on multicanonical
Monte Carlo is proposed
for searching  rare trajectories in chaos.
The proposed method is
tested with four-dimensional coupled standard maps
and  successfully computes the probability
of  the forgetting time $T_\epsilon$ 
down to $\sim 10^{-14}$.

Applications of the proposed method 
to dissipative and/or multi-basin systems
will  be interesting, as well as  
search for highly unstable trajectories in weakly chaotic
systems. Another interesting subject is to develop a way
to relate $P(T_\epsilon)$ to ``escape rate''
and hence interpret it in the thermodynamic 
formalism~\cite{BeckSchloegl199501}, which connects
our approach to existing studies on large deviations
in chaos. 

\section*{Acknowledgements}
The authors would like to thank M. Kikuchi  and T.~Chawanya for their
helpful comments.
This work was supported in part by Global COE Program (Core Research
and Engineering of Advanced Materials-Interdisciplinary Education
Center for Materials Science), MEXT, Japan.
Numerical experiments are carried out on PC
clusters at Cybermedia Center, Osaka University.

\bibliographystyle{cpc}
\bibliography{CCP2009}

\end{document}